\documentclass[12pt]{article}
\pdfoutput=1
\usepackage[utf8]{inputenc}
\usepackage{amsthm}

\usepackage{rotating, multirow}
\usepackage{titlesec}
\usepackage{titling}
\usepackage{fullpage}
\usepackage{graphicx,color}
\usepackage{amsfonts}
\usepackage[english]{babel}
\usepackage{multirow}
\usepackage{mathtools}
\usepackage{helvet}
\usepackage[T1]{fontenc}
\usepackage{amsmath,amssymb}
\usepackage{mathrsfs}
\usepackage{amsthm}
\usepackage{alltt,verbatim,mathalfa}
\usepackage{tabularx}
\usepackage{soul}
\usepackage{algorithm}
\usepackage{hyperref}

\usepackage{authblk}
\usepackage{geometry}
\usepackage{graphicx}  
\usepackage{subcaption}
\usepackage{amsmath}
\usepackage{multirow}
\usepackage{adjustbox}
\usepackage{verbatim}
\title{Mapping the Russian Internet Troll Network on Twitter using a Predictive Model}%%

\author[1]{Sachith Dassanayaka}
\author[2]{Ori Swed}
\author[3]{Dimitri Volchenkov}

\affil[1,3]{Department of Mathematics and Statistics, Texas Tech  University, Lubbock,
Texas}
\affil[2]{Department of Sociology, Anthropology, and Social Work, Texas Tech  University, Lubbock,
Texas}

\date{}

\begin{document}

%%%%%%%%%%%%%%%%%%%%%%%%%%%%%%%%%%%%%%%%%%%%%%%%%%%%%%%%%%%%%%
\maketitle

\begin{abstract}
Russian Internet Trolls use fake personas to spread disinformation through multiple social media streams. Given the increased frequency of this threat across social media platforms, understanding those operations is paramount in combating their influence. Building on existing scholarship on the inner functions within influence networks on social media, we suggest a new approach to map those types of operations. Using Twitter content identified as part of the Russian influence network, we created a predictive model to map the network operations. We classify accounts type based on their authenticity function for a sub-sample of accounts by introducing logical categories and training a predictive model to identify similar behavior patterns across the network. Our model attains 88\% prediction accuracy for the test set. Validation is done by comparing the similarities with the 3 million Russian troll tweets dataset. The result indicates a 90.7\% similarity between the two datasets. Furthermore, we compare our model predictions’ on a Russian tweets dataset, and the results state that there is 90.5\% correspondence between the predictions and the actual categories. The prediction and validation results suggest that our predictive model can assist with mapping the actors in such networks.

 \emph{keywords: Predictive Model, Authenticity Function, Validation }

\end{abstract}

\section{Introduction}

Internet troll networks (ITN) or troll networks—entities that spread disinformation on social media through fake accounts—have become a serious threat to democratic discourse. Understanding and detecting those operations turns paramount. One of the leading state actors in this field was Russia \cite{ref25}, using ITN across multiple platforms and in different countries as an influence tool\cite{ref36,ref56}. Examination of the Russian trolling operation indicates that Russian ITN was fragmented and run by multiple entities. One of those was the Internet Research Agency (IRA), a company affiliated with the Kremlin and located in St. Petersburg. The IRA has been involved with multiple overseas political activities, among them the 2016 U.S. presidential election interference \cite{ref36,ref55,ref24}. In the 2016 campaign, the IRA accounts concealed their true identity by presenting themselves as legitimate social media users such as regular folks from rural and suburban America \cite{ref37}. This common tactic makes it harder for users and regulators to identify and prevent ITNs and influence operations. Thus, the problem is differentiating fake actors that pretend to be legitimate users from the rest. Due to the hidden nature of the Russian ITN, differentiating actors based on their activities is an open challenge. Furthermore, researchers’ difficulties with identifying the fake users pose a significant challenge for understanding those types of operations and consequently make it harder for us to devise ways to stop them. To address this problem, in this study, we use artificial intelligence as an ancillary instrument to map trolls’ activities across influence networks.

The principal obstacle scholars that study ITN face is the amorphous nature of the network. Paradoxically, the personae that the network introduces maintain a legible and influential social appearance. However, these personae are fake and, as such, offer a limited foothold on the perpetrators. Moreover, they are hard to detect, differentiating between authentic users and socket puppets. Following existing methods in this field, our study addresses this obstacle by studying exposed trolls’ networks, suggesting a machine learning predictive model to map network operations and classify the actors. The predictive model focuses on different types of actors in the Russian ITN, their distribution, and their activities. Building on studies that identified organizational order and functions assigned to specific accounts, we assert that this division of roles is meaningful enough to help with mapping those networks \cite{ref2,ref68}. 

We used English language tweets from the IRA tweets dataset that was linked with the IRA, from the Alliance for Security Democracy data (\url{https://securingdemocracy.gmfus.org/}) and mapped them according to their authenticity function. In other words, the different ways they presented themselves. We recognized four conceptual categories: \textit{Fake News, Organizations, Political Affiliates,} and \textit{Individuals}. Next, we trained a machine learning predictive model to identify the category of each account in the network using several features, such as the \textit{number of tweets, retweets,} and \textit{followers}. One-third of the English language dataset data were partially hashed, which means there was no available information on the account in its description and name to assert into which categories they fit. As such, we used our predictive model on one-third of hashed data. Moreover, we validated our predictive model with the 3 million Russian Troll tweets dataset \cite{ref2}. Russian speakers assisted us with manual coding of the accounts per the four categories on the Russian language dataset, a subset of the IRA tweets dataset, to further evaluate the predictive model's accuracy.

\section{Background and Related Work}

Russian ITN on social media has received much attention from scholars and practitioners attempting to better understand this trend and phenomenon. The 2016 Russian influence troll campaign on social media was excessively studied by cyber security experts and scholars. This event reoriented discussion on ITN and brought much attention to Russia as a perpetrator of influence campaigns in social media and the 2016 elections as a case study \cite{ref36}. Focusing on the IRA operation between 2015 to 2017, Boatwright, Linvill, and Warren (2018) explored how the organization weaponized social media and spread its agenda and messaging across multiple platforms \cite{ref2}. The researchers shed light on the network, from the way it created a network of fake personae, organizations, and websites to the way that it spread messaging in fabricated echo chambers. They recognized five types of Twitter actors whose behaviors were drastically dissimilar: \textit{right troll, left troll, news feed, hashtag gamer,} and \textit{fearmonger} based on the IRA. In the same context, they pointed out that the Russian intervention in overseas political activities through fake personae exacerbated social drift by misusing the social media streams such as Twitter. Another study, which replicated Boatwright, Linvill, and Warren’s (2018) research concerning category mapping, was conducted by Lewinski and Hasan \cite{ref65}. The researchers attempted to replicate the original categories in the 3 million Russian Troll Tweets dataset and automate the classification using a machine learning model. 

Elaborating on their earlier work, Linvill et al. (2019) examined tweets connected with the IRA, creating a more comprehensive categorization of behaviors within the network related to the 2016 U.S. Presidential election \cite{ref70}. They tested tweets that were released by the ITN a month before the election and examined the communication distinctions among Twitter actors. They came up with seven behavioral categories actors produce in the network: attack left, support right, attack right, support left, attack media, attack civil institutions, and camouflage. 

The scholarship on ITN went beyond theoretical and conceptual discussions, focusing on detection. Methods of detection on Twitter were frequently using a predictive model that focused on specific aspects of the actors or their behavior, for example, the political role of Internet trolls (Atanasov et al. 2019) \cite{ref31}. The researchers examined behavior patterns of political trolls on the IRA Russian Troll dataset and automated the classification using supervised, and distance supervised learning machine learning methods to treat categorized and non-categorized trolls, respectively. Bidirectional Encoding Representations from Transformers (BERT) was used to build a machine learning model to determine the political tendency of trolls with the Russian Troll dataset by Chun et al. \cite{ref6}. Kim et al. (2019)\cite{ref33} suggested a text distance metric (i.e., a time-sensitive semantic edit distance) and applied it to classify the Russian trolls. 

Although several studies have addressed the importance of Internet troll classification in the literature, due to the artificially generated nature of the Russian ITN, differentiating the actors and revealing the structure still pose a challenge. Some common issues, such as abstract categorization based on actors’ behavior, automating the operation of this categorization, and testing the proposed solution on a publicly available troll dataset, are still open to scholars.

\section{Dataset Description}

This study made use of a publicly available dataset of identified troll activity in social media. The data, which covers multiple actors and operations, is presented by the Alliance for Security Democracy. Our study focused on a particular actor, operation, and platform. Namely, the IRA operation of Twitter in the period leading to the 2016 intervention operation in the US elections and the time-frame right after \cite{ref24,ref35}. The dataset consisted of about nine million tweets using 58 languages, and some information had been hashed, i. e., some information such as ``user profile description'' was unavailable due to the privacy concerns by Twitter. We recognized about forty features in the dataset, including profile-related features, qualitative behavioral measures, and tweet-related linguistic features. We extracted the English language tweets (henceforth: $1^{st}$ dataset) and Russian language tweets (henceforth: $2^{nd}$ dataset), which were about three million and four million tweets, respectively. There were 2,832 unique Twitter actors that linked with the IRA English dataset, and the tweets were posted between November 2009 and May 2018. Out of those, about 36\% of the accounts were hashed, including nearly 21\% accounts that had at least one hashtag. The rest of the 15\% hashed accounts were not included in any hashtags. Since we used a machine learning predictive model to recognize the accounts based on their activities, we had to validate our model to ensure it performed well beyond the training data. Therefore, for the validation purpose, we used the 3 million Russian Troll tweets dataset (henceforth: 3$^{rd}$ dataset) and the $2^{nd}$ dataset. The $3^{rd}$ dataset includes 1,133 IRA-related actors, while the $2^{nd}$ dataset consisted of 1,554 actors in total, but we removed 119 accounts due to lack of information, and we used the remaining 1,435 actors from the $2^{nd}$ dataset to validate the predictive model.  

\section{Approach}

In this section, we report the hypothetical categories and feature selection of our study. We created a unique filter for the actors in the Russian IRA network by introducing a conceptual categorization of troll actors. The user profile description feature carried more valuable information regarding the actors' interest and their unique backgrounds. Across all cases, the Russian trolls attempted to promote directly or indirectly a specific agenda while maintaining the façade of a real user (not a ``sack puppet''). Keeping that façade, which allowed them to maintain authenticity, forced the users to specific patterns of behavior and specifically identifiable categorizations. To identify those actors and describe their behaviors within the Russian ITN, we introduced four conceptual categories, which based on the fake authenticity of actors:

\begin{itemize}
%\begin{enumerate}[label=\Roman*.]
	\item Fake News: Accounts whose description was of a news outlet, including government news, private news organization, and specialty news such as UFO, military, regional, city, etc. Some of the actors introduced themselves to the network as online news sources, such as \textit{San Jose Daily, San Francisco Daily, Novosibirsk Bulletin,} and \textit{Memphis Online}, to imply that they provided updates by connecting with authorized local news sources and tweets regarding local news trends. These actors' portrayals benefited the image of original news mediums rather than a news fan or an individual that shared the news on a specific subject. This description differed from the News Feeds category of Linvill and Warrens' (2018) research study. Under the News Feeds category, some actors in the network were pretending to be individuals who talk political interests, such as \textit{TENGOP, Political Observer,} and \textit{Special Affair}.  
	
	\item Organizations: Accounts whose description was of non-governmental organizations or businesses, including volunteer organizations. News organizations were not included in that category. Some Twitter actors let others imagine that they were a group of people who could support the community, such as \textit{Heart Of Texas} and \textit{Black To Live,} pretending to be social movements or commercial entities.
	
	\item Political Affiliates: Accounts whose description was politics-related or seems to present as individuals that are overtly politically affiliated. Here we could find accounts such as Jenna Abram that provide unfiltered political commentary.
	
	\item Individuals: Accounts whose description was of individuals or seems to present as individuals that are not overtly politically affiliated. Lastly, the ITN is full of ``regular folks'' that function as the core of the network, directing real users into the network and echoing its messages.
	
%\end{enumerate}	
\end{itemize}
We extracted features such as \textit{user id, user name, user profile description, user mentioned count, tweet count, retweet count, followers, followings, reply,} and \textit{likes} from the $1^{st}$ dataset and associating them with each category, using them as the indicators for the actors’ behavior. We applied the authenticity function on the \textit{user profile description} feature to assign categories to the actors in this social network. Although we performed a manual categorization based on \textit{user profile description} for about eighteen hundred actors, nearly a thousand of them did not have this feature available due to hashed data. Thus, we had to find a way to treat those actors who had a hashed \textit{user profile description} by proposing a mathematical approach approximating the corresponding categories.

\subsubsection{Relative frequency of actors' categories}

\begin{figure}[h!]
	\centering
	\centerline{\includegraphics[scale=0.7]{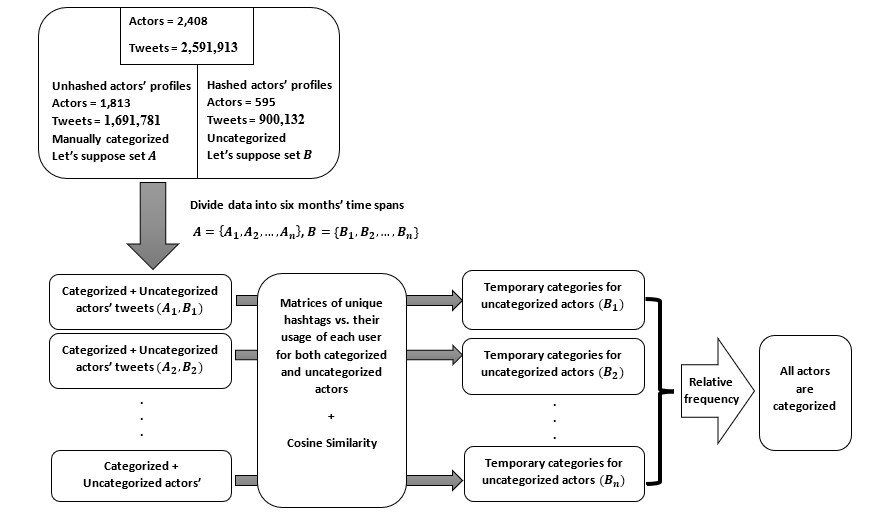}}
	\caption{Process of identifying categories for 21\% hashed actors in the $1^{st}$ dataset according to the four conceptual categories.}
	\label{fig:process}
\end{figure}

At this stage, we categorized about 65\% of the hashed accounts in the $1^{st}$ dataset using their \textit{user profile description}. This sample size was not sufficient for the training of the AI (see Appendix-A). To increase our sample, we used repeating hashtags that appeared in the identified and categorized accounts to match with similar clusters of hashtags in accounts that have no description (hashed accounts). By accounting for similarities in hashtags and using them as a key, we were able to add an additional 21\% of the sample. This addition improves the robustness of our model. Therefore, we used 85\% of the data from the dataset for the following stages of our study. There were two types of real-time complications that we had to face while we were analyzing hashtags. The first problem was dealing with text or human languages in order to process hashtags. The second problem was that the usage of a hashtag could be changed with time; in other words, a particular hashtag became more prevalent during an election and then disappeared. We treated those complications simultaneously by using open-source libraries and taking subsamples with fixed time frames. Natural language processing (NLP) is the key to dealing with human languages as we used the natural language toolkit (NLTK) platform in Python, which has over 50 corpora and lexical supports, to process tweets and hashtags \cite{ref8}. We used these libraries to analyze the data in individual subsets from July 2009 to June 2018, with each subset representing six months of fixed time in that span (see Fig. \ref{fig:process}). Each subset was analyzed separately. It was an important task to clean some unnecessary symbols and convert all of the hashtags to either capital or simple letters. This prevented the capitalization and symbolical errors that may occur during the analysis and was done through the NLTK package.

First, the initial subspan and all of the unique hashtags used by both groups of actors in the first span were compiled into a list, and this list was used to create a collection of vectors for each actor. To form the vector, we cross-referenced every actor in both the unhashed (categorized) and hashed (uncategorized) groupings with each of the unique hashtags. Each row represented a unique hashtag, and if the actor used a given hashtag, it was represented with a 1 (true) or a 0 (false).

Consider the collection of vectors of uncategorized actors denoted by $ U $, as in:

\begin{equation}\label{eq1}
U = 
\left\{ u_{.1} , u_{.2} , \cdots , u_{.s} \right\}\\ 
=
\left\{ 
\left[ \begin{array}{c}
     u_{11}\\
     u_{21}\\
     \vdots\\
     u_{m1}
\end{array}\right] \\ 
,
 \left[\begin{array}{c}
     u_{12} \\
     u_{22} \\
     \vdots \\
     u_{m2}
\end{array}\right] \\
,
\cdots
,
 \left[\begin{array}{c}
     u_{1s} \\
     u_{2s} \\
     \vdots \\
     u_{ms}
\end{array}\right] \\
\right\},\\ 
\end{equation}

where $m$, $s$ represent the total number of unique hashtags, and number of hashed actors in the current time span, respectively. In the same fashion, the  collection of the vectors of categorized actors in the same span denoted by $ V $, as in:

\begin{equation}\label{eq3}
V = 
%\begin{pmatrix}
\left\{ v_{.1} , v_{.2} , \cdots , v_{.k} \right\}\\ 
%\end{pmatrix}
=
\left\{ 
\left[ \begin{array}{c}
     v_{11}\\
     v_{21}\\
     \vdots\\
     v_{m1}
\end{array}\right] \\ 
,
 \left[\begin{array}{c}
     v_{12} \\
     v_{22} \\
     \vdots \\
     v_{m2}
\end{array}\right] \\
,
\cdots
,
 \left[\begin{array}{c}
     v_{1k} \\
     v_{2k} \\
     \vdots \\
     v_{mk}
\end{array}\right] \\
\right\},\\ 
\end{equation}

where $k$ represents the total number of unhashed actors in the current time span.\\

{Algorithm:}\\

     \begin{itemize}
    \item Let $i \in s$ and $j \in k$
    \begin{itemize}
    
    \item Now, consider the current span
    \item From $i=1$ (until $i=s$)
        \begin{itemize}
        \item Next, we can calculate the cosine similarity value between $u_{.i}$'s and $v_{.j}$'s via, 

        \begin{equation}\label{eq4}
        CS(u_{.i},v_{.j}) = \frac{u_{.i}.v_{.j}}{||u_{.i}|| \;\; ||v_{.j}||} ,
        \end{equation}

            where $ i \leq s$ and $ j \leq k$ \\
            \item find $max \; CS(u_{.i},v_{.j}) $  
            \item $category(u_{.i})=category(v_{.j})$
            \item $i=i+1$
        \end{itemize}
    \item Repeat above steps for the next span
    
    \end{itemize}
    \item Finally, consider the relative frequency distribution of each actor in uncategorized set across all spans $(\in B)$ to approximate their abstract category.
    \end{itemize}
    
Per the algorithm, we assigned an impermanent category to $u_{.i}$ of the corresponding actor $v_{.j}$ by considering $max(CS(u_{.i}, v_{.j}))$ \cite{ref10,ref40}. We continued this process until the categories for all the $u_{.i}$'s (hashed actors) were approximated in the current subset. This process was recursive, continuing until determining the categories of all hashed actors in all the subsets. Finally, we listed the relative frequency distributions for all hashed actors regarding their impermanent categories across the subsets and the highest frequent category assigned to the particular actor. No more than one mode (the highest frequent categories) appeared in each relative frequency distribution during the calculation. This analytical approximation assisted us in approximating categories for all of the hashed actors in this dataset, and then we could implement a model that can be used to recognize actors.
  
\begin{figure}[h!]
	\centering
	\centerline{\includegraphics[scale=1]{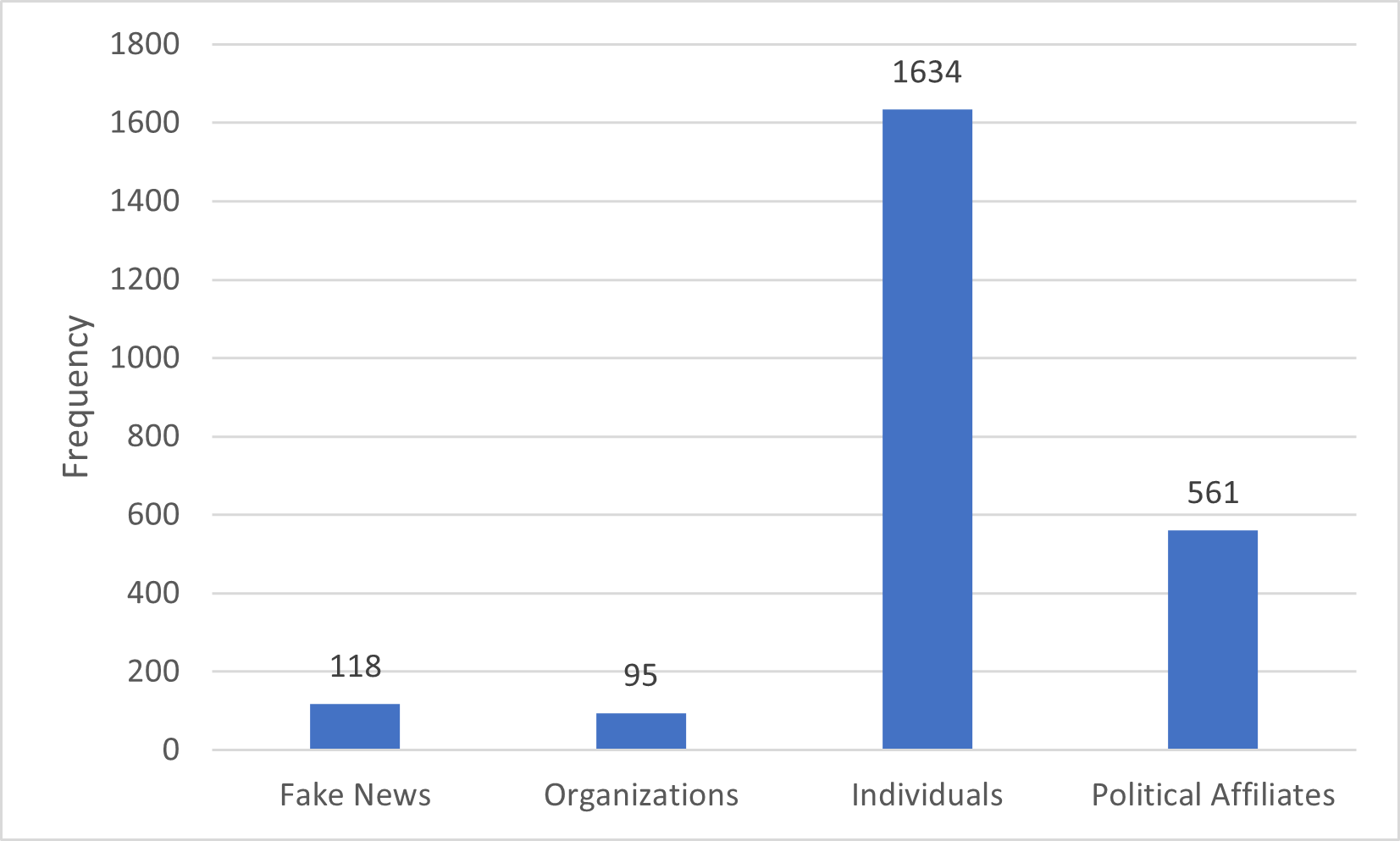}}
	\caption{Frequency distribution of abstract categories for 2,408 actors in IRA English dataset.}
	\label{fig:frequencyDis}
\end{figure}

After the categorization, the frequency distribution of conceptual categories illustrated that the IRA English tweets dataset was imbalanced due to some categories having fewer observations than the other categories \cite{ref11,ref12,ref13} (see Fig. \ref{fig:frequencyDis}).

\subsection{Feature Selection}

After completing the approximation of categories for the 21\% of the hashed actors, we recognized about 85\% of the actors in the IRA English dataset, fitting with the four conceptual categories. At the following stage, with 2,408 categorized actors (85\%), we used features that are associated with each account to enhance accuracy and reduce over-fitting and training time. The IRA English dataset had more than 30 features (see Appendix-B for the list of features in the $1^{st}$ dataset). Not all features demonstrate meaningful relations with our categorization, and consequently, we dropped them out of the model. We determine the strength of these features by using a statistical-based feature selection criterion, the ``SelectKBest'' class in the scikit-learn library, for our univariate feature selection criterion in Python. The ``SelectKBest'' algorithm uses the chi-square test as the scoring function sequential procedure, which measures the connection between two categorical features, and ranks the features according to their importance \cite{ref16,ref50}.

\begin{table}[t!]
\centering
\caption{\rm Ranking important features of IRA English dataset using the ``SelectKBest'' class in scikit-learn library.}
\setlength{\tabcolsep}{20pt}
\footnotesize{
\begin{tabular*}{8cm}{cc}
\hline\hline
\raisebox{-2ex}[0pt][0pt]{Features} &  \raisebox{-2ex}[0pt][0pt]{Score} \\
 \\\hline

    followers count & 9867.6 \\ 
     hashtags count & 9619.8\\ 
      tweets count & 2739.45\\ 
   usersMentioned & 2269.14\\ 
       retweets count & 1448.3\\ 
           followings count  & 1324.99\\ 
        likes count &  1262.93\\ 
         replies count  & 876.12\\ 
        quotes count  & 9.3207\\ 
        urls count & 2.9191\\ 
        polls count & 2.7328\\ 
   retweet tweet ratio &  1.4898\\ 
   $\cdots$            &  $\cdots$\\ 
    retweeters category &  0.12836\\ 

\\\hline\hline\end{tabular*}}
\label{tblFeatures}
\end{table}

The selected features are \textit{tweets count, retweets count, followers count, followings count, replies count, likes count, users Mentioned,} and \textit{hashtag counts} (see Table. \ref{tblFeatures}). Furthermore, we check Pearson’s correlation coefficient of the selected eight features \cite{ref28}. The selected features reduce the multicollinearity, which is the occurrence of high correlations among two or more independent variables in the dataset (see Appendix-C). To avoid the effect of different scales across features, we normalized the dataset. For example, the \textit{followers count} could be thousands for a particular actor, but \textit{user Mentioned} count was less than ten. Therefore, the scales for these two features are different, so we reduced the effect of this scale difference by normalizing (L1) the set. Next, we applied supervised machine learning techniques to build a predictive model by considering 85\% of accounts (2,408) from the $^{1st}$ dataset. Normalization was not enough since there was still an imbalance in the sample that can affect the supervised learning classification \cite{ref57,ref58}. Our imbalance dataset consisted of majority and minority categories that had a larger number of samples and a smaller number of samples, respectively. Thus, there was a higher chance of miscategorizing the minority classes than the majority classes when training the predictive model, which could impede performance \cite{ref58}. To address that, we used stratified subsampling and Bootstrapping technique \cite{ref12}. 

We used three traditional measurements precision, recall, and \textit{f1}-score to measure the predictability of the model \cite{ref22,ref23}. 

\begin{figure}[h!]
	\centering
	\centerline{\includegraphics[scale=1.32]{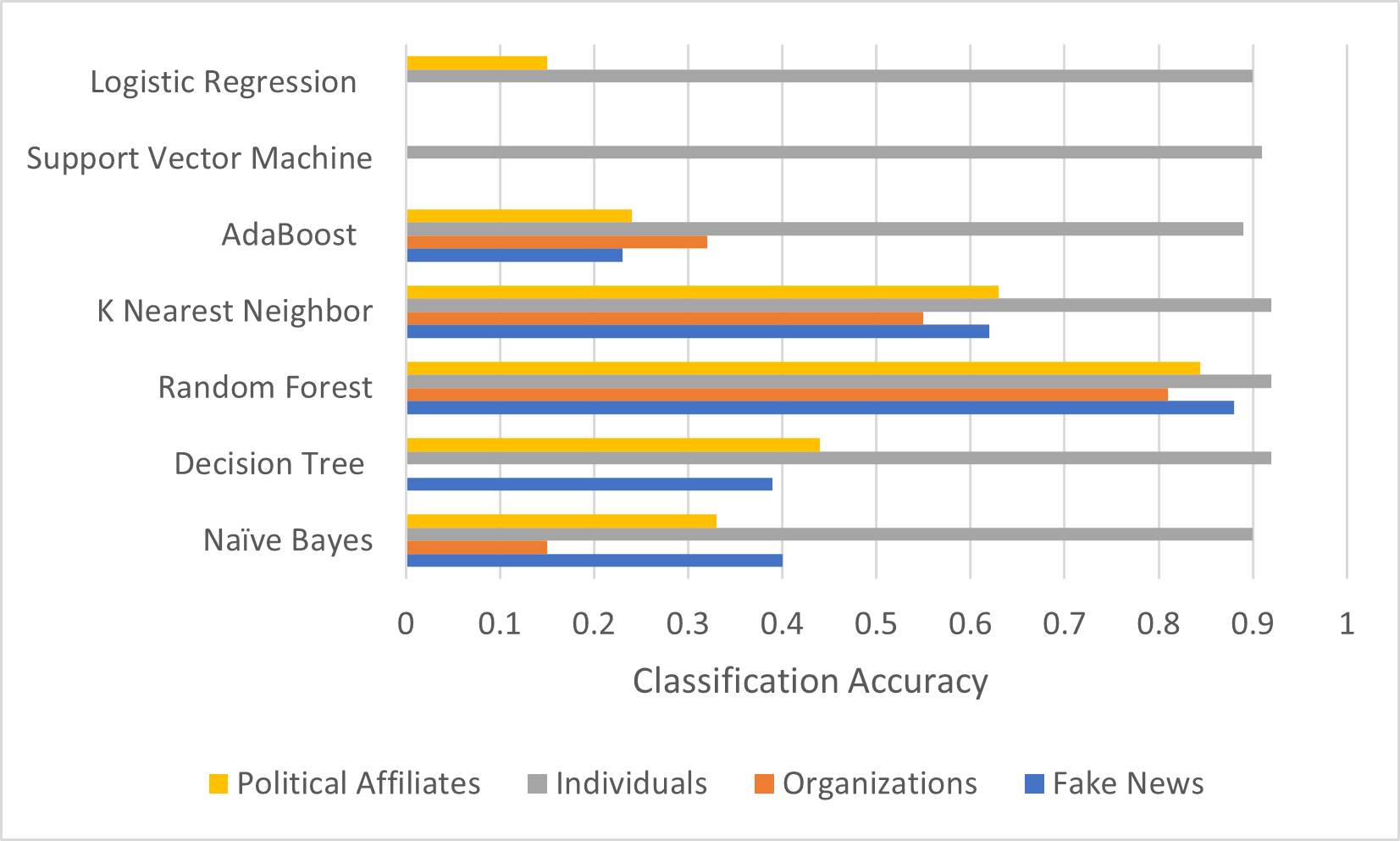}}
	\caption{Classification accuracy for each conceptual category under different classifiers with the selected eight features.}
	\label{fig:classiCompare}
\end{figure}

Given that our metric of success was classification accuracy, we used a classifier that also provided a solution for the imbalance in the sample. We used seven frequently use classifiers (Naïve Bayes, Support Vector Machine, Adaboost, K Nearest Neighbor, Random Forest, Decision Tree, Logistic Regression) to compare overall predictability, \cite{ref15,ref60,ref61,ref62,ref63,ref41}. A comparison of the classifiers showed variation in accuracy scores (see Fig. \ref{fig:classiCompare}). The Naïve Bayes classifier was 40\%, 15\%, 90\%, and 33\% accuracy for Fake News, Organizations, Individuals, and Political Affiliates, respectively, while the Decision Tree classifier provided 39\%, 0\%, 92\%, and 44\% accuracy for the same abstract categories (see Table 2). Further, although Logistic Regression and Support Vector Machine could classify the Individuals category with about 90\% accuracy, both Fake News and Organizations categories have 0\% support while Logistic Regression indicated 15\% accuracy for Fake News abstract category. On the other hand, AdaBoost, K Nearest Neighbour, and Random Forest classifiers were 23\%, 62\%, 88\% for the Fake News and 32\%, 55\%, 81\% for Organizations, respectively. The Individual category had the highest accuracy of 91\%, 89\%, 92\% while Political Affiliates indicated 24\%, 63\%, 84\% for AdaBoost, K Nearest Neighbour, and Random Forest classifiers separately. Thus, the Random Forest classifier provided greater accuracy for each abstract category.

\begin{table}[t!]
\centering
\caption{\rm Overall classification accuracy for different classifiers.}
\setlength{\tabcolsep}{20pt}
\footnotesize{
\begin{tabular*}{10cm}{cc}
\hline\hline
\raisebox{-2ex}[0pt][0pt]{Classifier} & \raisebox{-2ex}[0pt][0pt]{Weighted average f1-score} \\  
\\\hline
Naïve Bayes                                   & 81.67                     \\ 
Decision Tree                                 & 85.18                     \\ 
Random Forest                         & 88.15                     \\
K Nearest Neighbor                   & 85.93                     \\ 
AdaBoost                                      & 79.26                     \\ 
Support Vector Machine                  & 84.04                     \\ 
Logistic Regression                           & 82.22                     \\ 
\hline
\end{tabular*}}
\label{tbl2}
\end{table}

\begin{figure}[h!]
	\centering
	\centerline{\includegraphics[scale=0.8]{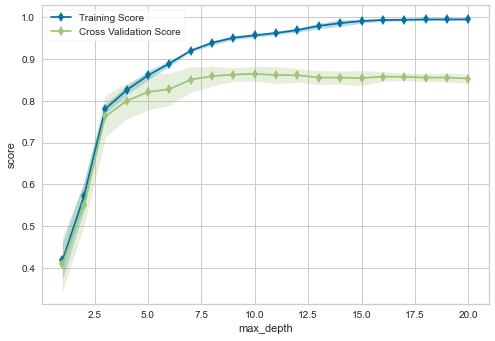}}
	\caption{Accuracy score (average \textit{f1}-score) of the forest changes with the average depth of the trees.}
	\label{fig:treeDepth}
\end{figure}

Since Random Forest Classifier is a straightforward machine learning algorithm that learns imbalanced data, we used this cost-sensitive learner to develop our predictive model \cite{ref15,ref41}. The random forest is a collection of many decision trees that keeps the minimum relationship among trees. In this forest, we could see the performance, in other words, the predictability, change with the depth of the trees (see Fig. \ref{fig:treeDepth}). Controlling the optimum depth of the tree was significant because too few levels lead the tree to under-fit, and a higher number of levels prone the model to over-fit to the training samples. Our forest, theoretically, provided a higher accuracy for five average trees depth without under-fit or over-fit.

The Scikit-learn library in Python supports Random Forest Classifier that allowed us to control the subsample through the parameter \textit{class weight} set equals to \textit{balanced subsample} \cite{ref16}. The \textit{balanced subsample} mode allowed the classifier to regulate weights automatically according to the inverse proportional to the class frequency with data based on bootstrap samples \cite{ref12}. In return, the effect of imbalanced data was reduced. Furthermore, through the Scikit-learn, we used Gini Index to measure the probability of features being wrongly classified and identify the root node of the decision trees \cite{ref20,ref21}. The Gini index values are bounded between 0 and 1, inclusive, while 0 represents the purity of the classification, and one expresses the randomness. Let us suppose that there are $ k \in \mathbb{Z^{+}}$ classes in the dataset, and the Gini impurity can be defined via,

 \begin{equation}\label{eq5}
 G(S)= \sum_{i=1}^{k} P_{i}(1-P_{i}),
 \end{equation}

where $S$ denotes the dataset and $P_{i}$ represents the probability of selecting a observation from $i^{th}$ class. The Random Forest Classifier controls the greatest possible depth of each decision tree by expanding nodes until each branch of a tree classifies into a unique category. At this point, we built our predictive model and tested the predictability of the model using traditional measurements results based on the precision, recall, and \textit{f1}-score on the test set, which illustrates the model accuracy, not only the overall but the individual categories as well (see Table. \ref{tblMeasurmnts}).

\begin{table}[t!]
\centering
\caption{\rm Predictability measures for each class based on precision, recall, and f1-score on the test dataset.}
\setlength{\tabcolsep}{20pt}
\footnotesize{
\begin{tabular*}{12cm}{cccc}
\hline\hline
\raisebox{-2ex}[0pt][0pt]{Category} &  \raisebox{-2ex}[0pt][0pt]{precision}    &  \raisebox{-2ex}[0pt][0pt]{recall}  &  \raisebox{-2ex}[0pt][0pt]{f1-score}\\
& &  &  \\\hline
Fake News            & 0.94               & 0.82            & 0.88      \\
Organizations        & 0.82               & 0.80            & 0.81   \\
Individuals          & 0.88               & 0.96            & 0.92   \\
Political Affiliates & 0.83               & 0.85            & 0.84  
\\\hline\hline\end{tabular*}}
\label{tblMeasurmnts}
\end{table}

After the stratified 5-fold cross-validation, the result was 88\% of predictability accuracy on the test set.

\subsection{Predictive model validation}

\begin{figure}[h!]
	\centering
	\centerline{\includegraphics[scale=1]{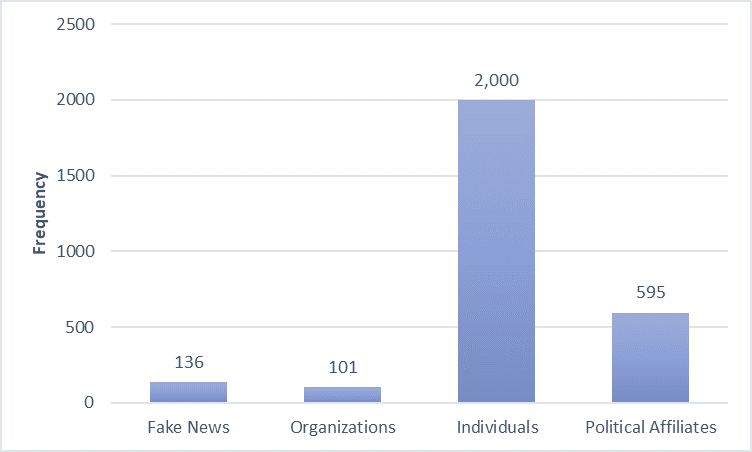}}
	\caption{Frequency distribution of conceptual categories for 2,832 actors in the 1st dataset, including all hashed accounts.}
	\label{fig:allcatfrqnz}
\end{figure}

Before we applied our predictive model to a random dataset and differentiated the actors of the particular network, the best practice was to validate the model's accuracy with known data. We performed the stratified fivefold cross-validation on the test set, reaching 88\% accuracy. For validation of this prediction, we tested the model performance on other datasets. To assure maximum compatibility in our first validation test, which used a similar dataset as we did, we used our prediction model to categorize the missing 15\% (about four hundred) hashed accounts that we could not either identify manually or match with hashtags in the $1^{st}$ dataset. This increased our sample to 2,832 accounts (see Fig. \ref{fig:allcatfrqnz}) and improved the compatibility with the Linvill and Warren database, which we used for our first validation test. 

\subsubsection{The first validation: with $3^{rd}$ dataset}

The 3 million Russian Troll Tweets dataset was published by FiveThirtyEight and Linvill, and Warren in 2018, based on their research study \cite{ref2}. In their study Linvill and Warren created a unique set of categorizations for different accounts, identifying them by functions. Our four conceptual categories do not fit with most of their functional classification (for example, left or right leaning trolls). The only similar comparable groups, and the ones we could use for the validation test, were the \textit{News Feed} from their study and \textit{Fake News} from our research study. Those were not fully compatible since Linvill and Warren’s \textit{News Feed} category included accounts that pretended to be individuals that specialized in sharing news. Yet, this specific category was similar enough, so it should include all the accounts we identified as \textit{Fake News}. This means that we could expect to see over 88\% of the accounts, our model identified as \textit{Fake News} and that shared with the Linvill and Warren data to be categorized under their \textit{News Feed} group. We filtered the unique actors who were labeled as \textit{Fake News} in our $1^{st}$ dataset and the actors who fitted into \textit{News Feed} group in the 3 million Russian Troll Tweets dataset as the second stage of our validation. We observed that there were forty-nine \textit{fake news} accounts that match with \textit{news feed} out of fifty-four handles. At the same time, five \textit{News Feed} handles were misclassified as \textit{individuals} by our predictive model. The result of this validation test suggested that our model prediction was 90.7\% accurate for \textit{Fake News} conceptual category. 

\subsubsection{The second validation: with $2^{nd}$ Dataset}

\begin{figure}[h!]
			\centering
			\centerline{\includegraphics[scale=1]{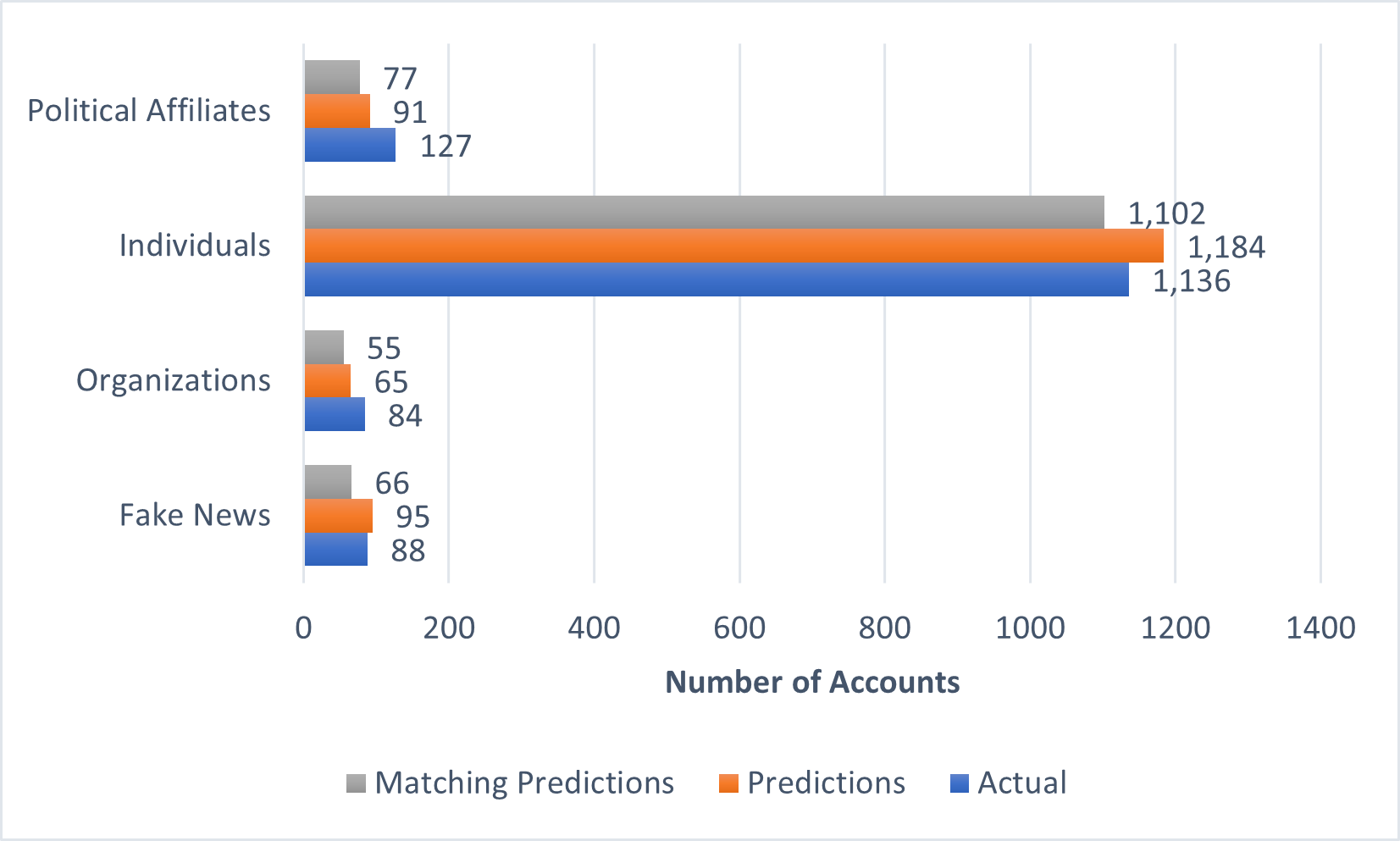}}
   			\caption{Comparison between number of actual and predicted accounts.}
   				\label{fig:valRussian}
\end{figure}

In a second validation test, we tested our model predictions with the  $2^{nd}$ dataset. Here we ran a similar process of identifying users and categorizing them manually. Next, we compared our manual categorization with the model's prediction. We assumed that our model would achieve the 88\% prediction level. Since we were working with the Russian language in the $2^{nd}$ dataset, we used the \textit{Google Translate Application Programming Interface(API)} to translate the Russian \textit{profile descriptions} to English. At the same time, Russian speakers verified the validity of 1,435 translated Russian language content and then assigned categories manually to the actors in the $2^{nd}$ dataset based on the same authenticity function. The manual categorization was used as a reference for the predictive model. We compared the manual categorization of actors in the four groups to the predictive model results on the dataset, testing for similarities between our predictions and the actual categories, namely, the manual categorization was done by the Russian speakers (see Fig. \ref{fig:valRussian}).

Fig.\ref{fig:valRussian} illustrates the comparison between the manual coding and the predictive model's results. It also captures the proportion of matching results. This validation test of the IRA Russian set shows that the predictive model achieves overall 90.5\% prediction accuracy, which is higher than anticipated.

\section{Discussion}

Russian trolls that linked with IRA spread out disinformation on Twitter by concealing their true nature and allowing them to engage in act of information warfare. Differentiate actors in such networks helped researchers to reveal and understand the structure of the troll networks and behaviors. 

In this study, we offered a mechanism that helps identify the actors in the Russian troll networks through a predictive model that has the ability to map the actors in the Russian troll network. Our model builds on four conceptual categories and assigns them to each actor based on their behaviors. The IRA dataset, which we used for our study, consists of about nine million tweets using 58 languages, including English and Russian. From this dataset, we used English tweets to propose conceptual categories and implemented a predictive model that provided 88\% prediction accuracy for the test set.

We verified our model predictions on the publicly available 3 million Russian troll tweets dataset, and the comparison result showed 90.7\% accuracy in prediction. Moreover, we validated the accuracy of our model predictions on the Russian language troll tweets set, which was a subset of the IRA dataset, by conducting a human evaluation. The outcome was 90.5\% similarity between the predictions and classification of the Russian language troll tweets dataset. Lewinski and Hasan attempted a similar analysis using a Support Vector Machine (SVM) model \cite{ref65}. With the SVM model, they tried to map Facebook's ads accounts while building on Linvill and Warren's categorization. Although their goal was not to focus on automating Linvill and Warren's categorization, their SVM model provided 75.5\% accuracy for the \textit{News Feed} category when Lewinski and Hasan applied to the 3 million Russian Troll Tweets dataset. The result demonstrated that our predictive model provided comparatively higher accurate predictions (90.7\%) than the SVM model to the \textit{News Feed} category in the 3 million Russian Troll Tweets dataset.

\begin{comment}

\section{Future Work}

In the future, Twitter activities such as tweets, retweets, likes, and replies will be used to examine activity patterns in the network and visualize them in 3-dimensional space. In order to find a Troll users' relationship among social media streams, the predictive model will be trained and validated on different social media platforms such as Instagram and Facebook. Mathematical models will be proposed to identify actors using these activity patterns. Further, we will use our \textit{tweet language free} network model for a few other networks on Twitter, including Chinese and Iranian, and analyze and compare these networks to reveal their typical patterns and behaviors. Although the research focuses on classifying fake personas, there is a high chance of improving this predictive model to identify real-time Troll activities on Twitter and differentiate them from legitimate users' activities. Further, to construct a real-time Troll activity tracking model, we will need to consider Twitter content as well in order to provide more accurate details to the model. Therefore, the system might be intelligent enough to isolate Twitter content and actions which belong to legitimate users from fake users. Furthermore, a Natural Language Procession and clustering technique will be required to analyze, compare, and group fake personas frequently to avoid loopholes in the Troll tracking system. 

\end{comment}

\section{Conclusion}

This study presents a path to reveal and study the structure of the IRA network; a hostile network that tried to promote a seditious agenda while concealing its activities and real identity. Since this study focuses on Twitter, the proposed solution (the predictive model) is based on troll networks that have been conducted on Twitter only. The organized nature of the tweets, either from a person or a computer, can be simplified using our conceptual categories, and the proposed model is a \textit{tweet language free} network model that is useful to map the IRA network in order to differentiate actors in ITN. Since the model is free of language, the process of classifying fake personas is not affected by the Twitter content's language as long as the samples are taken from the IRA network. We have to align with the IRA network because we have yet to test our predictive model against other networks outside the IRA.

Future research can build on this predictive model to better analyze and understand specifically the IRA network and influence operations online in general. Our model utilizes Twitter activities such as tweets, retweets, likes, and replies can be used to examine activity patterns in the network and visualize them in 3-dimensional space. Moreover, while the model was designed for Twitter, it can be reproduced for other social media platforms such as Instagram and Facebook. The predictive model offers us a stable baseline for further mathematical predictions and analysis of those actors and activities. Mathematical models that draw on our model can potentially identify actors. Further, this \textit{tweet language free} network model can be used to examine other influence networks on Twitter, among them the Chinese and Iranian networks, and analyze and compare these networks to reveal their typical patterns and behaviors. Although the research focuses on classifying fake personas, the foundations this study offers can potentially assist with improving existing tools that specialize in real-time identification of Troll activities on Twitter and differentiate them from legitimate users' activities.

\textbf{Appendix-A}

\begin{figure}[h!]
	\centering
	\centerline{\includegraphics[scale=1.2]{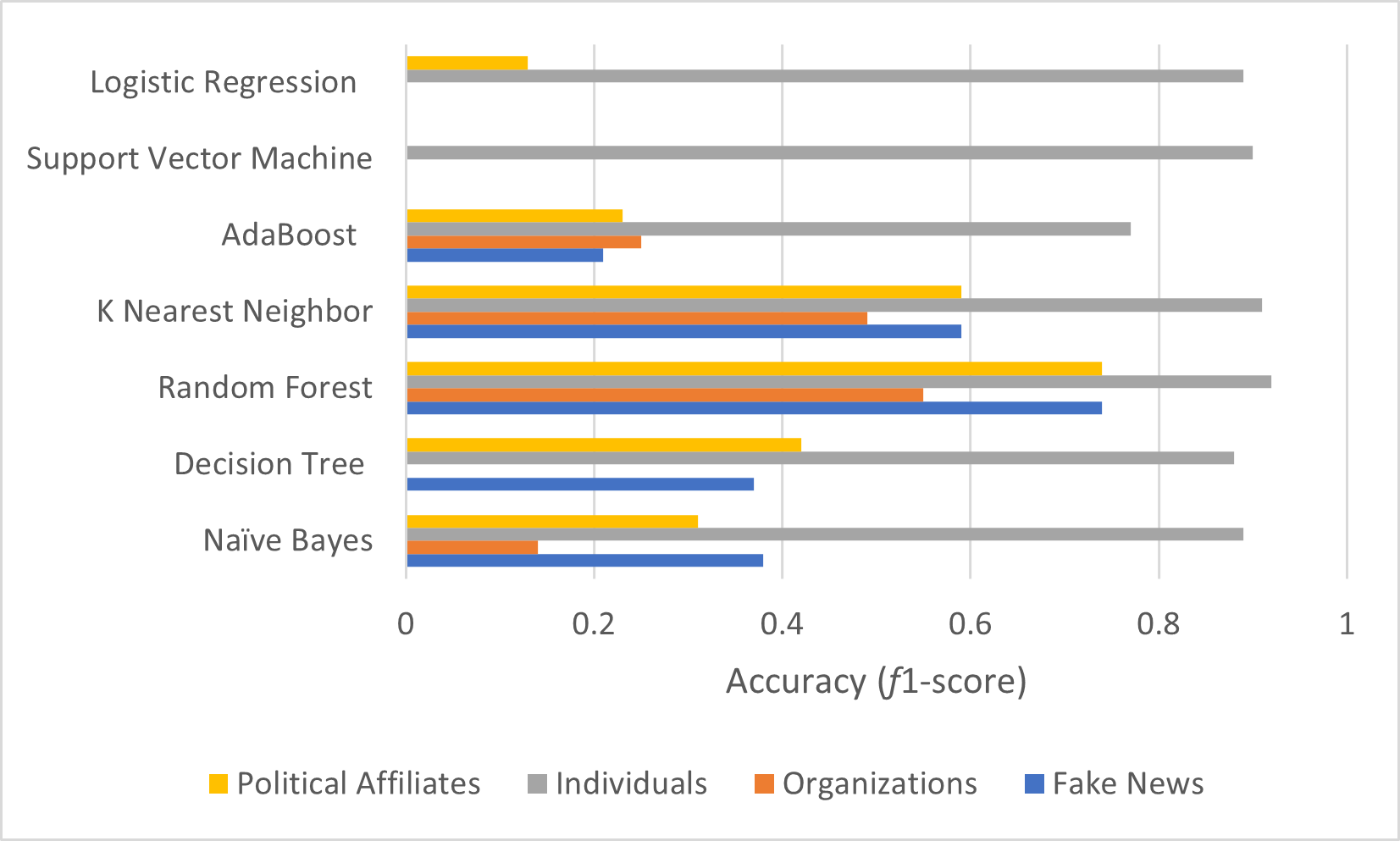}}
	\caption{Classification accuracy for each conceptual category under different classifiers with the selected eight features for unhashed accounts (64\%).}
	\label{fig:1800comparis}
\end{figure}

Suppose that we considered 64\% (about 1,800) actors in the 1st dataset and applied the above classifiers to the data in order to determine the ``the best'' classifier to train a predictive model. Classification accuracy was measured for different classifiers and conceptual categories (see Fig. \ref{fig:1800comparis}). According to Fig. \ref{fig:1800comparis}, regardless of the classifiers, accuracy declined, due to the sample size. Some categories did not have ``large enough'' samples, such as \textit{Organizations} to train the model. However, since the samples are small, models become biased, and the models predictability decreases.

The overall accuracy of the model decreases as accuracy for each category decreases. On the other hand, the majority of this unhashed data includes \textit{Individuals} actors, and it causes the misrepresents the overall accuracy. Therefore, considering hashed data as much as possible is the solution to overcome this issue. As a result, we used 21\% samples from hashed IRA English dataset to train our predictive model.

%\newpage
\textbf{Appendix-B}

\begin{table}[]
\centering
\caption{\rm The list of features in the $1^{st}$ dataset.}
\setlength{\tabcolsep}{20pt}
\footnotesize{
\begin{tabular*}{14cm}{cc}
\hline\hline
\raisebox{-2ex}[0pt][0pt]{Feature} & \raisebox{-2ex}[0pt][0pt]{Description} \\  
\\\hline
tweetid                    & Unique id for each tweet content                \\
userid                     & Unique id for each actor                        \\
user\_display\_name        & Actor name that public can see                  \\
user\_screen\_name       & Actor name on the account. \\
user\_reported\_location & Physical location of the actor when sending the tweet                          \\
user\_profile\_description & Profile description of the actor                \\
user\_profile\_url         & Profile url of the actor                        \\
follower\_count            & Number of followers related to the actor        \\
following\_count           & Number of followings related to the actor       \\
account\_creation\_date    & Account creation date                           \\
account\_language          & Language registered with the account            \\
tweet\_language            & Language of the tweet content                   \\
tweet\_text                & Tweet content                                   \\
tweet\_time                & Time of the tweet sent                          \\
tweet\_client\_name        & Name of the internet service provider           \\
in\_reply\_to\_tweetid     & Replied tweets id                               \\
in\_reply\_to\_userid      & Replied actors id                               \\
quoted\_tweet\_tweetid     & Quoted tweet tweet id                           \\
is\_retweet                & Whether tweet or retweet (True or False)        \\
retweet\_userid            & Retweet actor id                                \\
retweet\_tweetid           & Retweet tweet id                                \\
latitude                   & Latitude of the physical location of the actor  \\
longitude                  & Longitude of the physical location of the actor \\
quote\_count               & Number of quotes in the tweet                   \\
reply\_count               & Number of reported replies to the tweet         \\
like\_count                & Number of reported likes to the tweet           \\
retweet\_count             & Number of reported retweets to the tweet        \\
hashtags                   & Hashtags mentioned in the tweet                 \\
urls                       & Urls mentioned in the tweet                     \\
user\_mentions             & User ids mentioned in the tweet                 \\
poll\_choices              & Polls mentioned in the tweet                    \\
retweet tweet ratio      & Ratio between number of tweets and retweets                                    \\
usersMentioned             & Number of actors mentioned in the tweet         \\
hashtags count             & Number of hashtags mentioned in the tweet                    \\ 
\hline
\end{tabular*}}
\label{tbl2}
\end{table}

\newpage
\textbf{Appendix-C}

\begin{figure}[H]
	\centering
	\centerline{\includegraphics[scale=0.8]{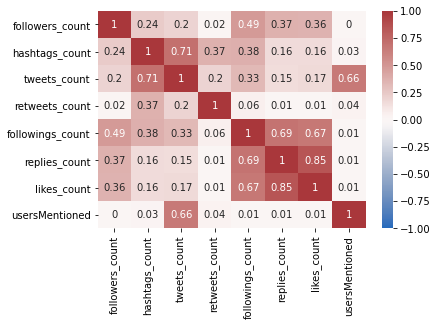}}
	\caption{Correlation matrix of the selected features.}
	\label{fig:corr}
\end{figure}

\end{document}